# Electric and Magnetic Dipoles in the Lorentz and Einstein-Laub Formulations of Classical Electrodynamics

Masud Mansuripur

College of Optical Sciences, The University of Arizona, Tucson, Arizona 85721

**Abstract**. The classical theory of electrodynamics cannot explain the existence and structure of electric and magnetic dipoles, yet it incorporates such dipoles into its fundamental equations, simply by postulating their existence and properties, just as it postulates the existence and properties of electric charges and currents. Maxwell's *macroscopic* equations are mathematically exact and self-consistent differential equations that relate the electromagnetic (EM) field to its sources, namely, electric charge-density $\rho_{\text{free}}$, electric current-density $J_{\text{free}}$, polarization $P$, and magnetization $M$. At the level of Maxwell's macroscopic equations, there is no need for *models* of electric and magnetic dipoles. For example, whether a magnetic dipole is an Amperian current-loop or a Gilbertian pair of north and south magnetic monopoles has no effect on the solution of Maxwell's equations.

Electromagnetic fields carry energy as well as linear and angular momenta, which they can exchange with material media—the seat of the sources of the EM field—thereby exerting force and torque on these media. In the Lorentz formulation of classical electrodynamics, the electric and magnetic fields, $E$ and $B$, exert forces and torques on electric charge and current distributions. An electric dipole is then modeled as a pair of electric charges on a stick (or spring), and a magnetic dipole is modeled as an Amperian current loop, so that the Lorentz force law can be applied to the corresponding (bound) charges and (bound) currents of these dipoles. In contrast, the Einstein-Laub formulation circumvents the need for specific models of the dipoles by simply providing a recipe for calculating the force- and torque-densities exerted by the $E$ and $H$ fields on charge, current, polarization and magnetization.

The two formulations, while similar in many respects, have significant differences. For example, in the Lorentz approach, the Poynting vector is $S_L = \mu_0^{-1} E \times B$, and the linear and angular momentum densities of the EM field are $\boldsymbol{p}_L = \varepsilon_0 E \times B$ and $\mathcal{L}_L = r \times \boldsymbol{p}_L$, whereas in the Einstein-Laub formulation the corresponding entities are $S_{EL} = E \times H$, $\boldsymbol{p}_{EL} = E \times H/c^2$, and $\mathcal{L}_{EL} = r \times \boldsymbol{p}_{EL}$. (Here $\mu_0$ and $\varepsilon_0$ are the permeability and permittivity of free space, $c$ is the speed of light in vacuum, $B = \mu_0 H + M$, and $r$ is the position vector.) Such differences can be reconciled by recognizing the need for the so-called *hidden energy* and *hidden momentum* associated with Amperian current loops of the Lorentz formalism. (Hidden entities of the sort do *not* arise in the Einstein-Laub treatment of magnetic dipoles.) Other differences arise from over-simplistic assumptions concerning the equivalence between free charges and currents on the one hand, and their bound counterparts on the other. A more nuanced treatment of EM force and torque densities exerted on polarization and magnetization in the Lorentz approach would help bridge the gap that superficially separates the two formulations.

Atoms and molecules may collide with each other and, in general, material constituents can exchange energy, momentum, and angular momentum via direct mechanical interactions. In the case of continuous media, elastic and hydrodynamic stresses, phenomenological forces such as those related to exchange coupling in ferromagnets, etc., subject small volumes of materials to external forces and torques. Such matter-matter interactions, although fundamentally EM in nature, are distinct from field-matter interactions in classical physics. Beyond the classical regime, however, the dichotomy that distinguishes the EM field from EM sources gets blurred. An electron's wave-function may overlap that of an atomic nucleus, thereby initiating a contact interaction between the magnetic dipole moments of the two particles. Or a neutron passing through a ferromagnetic material may give rise to scattering events involving overlaps between the wave-functions of the neutron and magnetic electrons. Such matter-matter interactions exert equal and opposite forces and/or torques on the colliding particles, and their observable effects often shed light on the nature of the particles involved. It is through such observations that the Amperian model of a magnetic dipole has come to gain prominence over the Gilbertian model. In situations involving overlapping particle wave-functions, it is imperative to take account of the particle-particle interaction energy when computing the scattering amplitudes. As far as *total* force and *total* torque on a given volume of material are concerned, such particle-particle interactions do not affect the outcome of calculations, since the mutual actions of the two (overlapping) particles cancel each other out. Both Lorentz and Einstein-Laub formalisms thus yield the same *total* force and *total* torque on a given volume—provided that hidden entities are properly removed. The Lorentz formalism, with its roots in the Amperian current-loop model, correctly predicts the interaction energy between two overlapping magnetic dipoles $m_1$ and $m_2$ as being proportional to $-m_1 \cdot m_2$. In contrast, the Einstein-Laub formalism, which is ignorant of such particle-particle interactions, needs to account for them separately.

**1. Introduction**. Maxwell's macroscopic equations relate the sources of the electromagnetic (EM) field, namely, free charge-density $\rho_{\text{free}}(\boldsymbol{r},t)$, free current-density $\boldsymbol{J}_{\text{free}}(\boldsymbol{r},t)$, polarization $\boldsymbol{P}(\boldsymbol{r},t)$, and magnetization $\boldsymbol{M}(\boldsymbol{r},t)$, to the field distribution throughout space and time $(\boldsymbol{r},t)$. The four fields appearing in these equations are the electric field $\boldsymbol{E}(\boldsymbol{r},t)$, the displacement $\boldsymbol{D}(\boldsymbol{r},t) = \varepsilon_0 \boldsymbol{E} + \boldsymbol{P}$, the magnetic field $\boldsymbol{H}(\boldsymbol{r},t)$, and the magnetic induction $\boldsymbol{B}(\boldsymbol{r},t) = \mu_0 \boldsymbol{H} + \boldsymbol{M}$. In the above expressions of $\boldsymbol{D}$ and $\boldsymbol{B}$, $\varepsilon_0$ and $\mu_0$ are the permittivity and permeability of free space. In their differential form, Maxwell's equations are written [1-6]:

$$\boldsymbol{\nabla} \cdot \boldsymbol{D} = \rho_{\text{free}}, \tag{1}$$

$$\boldsymbol{\nabla} \times \boldsymbol{H} = \boldsymbol{J}_{\text{free}} + \partial_t \boldsymbol{D}, \tag{2}$$

$$\boldsymbol{\nabla} \times \boldsymbol{E} = -\partial_t \boldsymbol{B}, \tag{3}$$

$$\boldsymbol{\nabla} \cdot \boldsymbol{B} = 0. \tag{4}$$

Let us forget for the moment the way by which these equations (which implicitly contain $\boldsymbol{P}$ and $\boldsymbol{M}$) were arrived at. Of course the Amperian current-loop model of a magnetic dipole, and the pair-of-charges-separated-by-a-spring model of an electric dipole, played important roles in arriving at these equations. However, ignoring the method of discovery of the equations and, instead, accepting them as natural laws, should make it clear that Eqs.(1)-(4) yield the field strengths $(\boldsymbol{E}, \boldsymbol{D}, \boldsymbol{H}, \boldsymbol{B})$ everywhere in space-time for given spatio-temporal distributions of $\rho_{\text{free}}$, $\boldsymbol{J}_{\text{free}}$, $\boldsymbol{P}$ and $\boldsymbol{M}$ *without* the need for any specific models of electric and magnetic dipoles.

There is nothing in the above equations to compel us to prefer one model of dipoles over the others, models that may be obtained quite simply by algebraic manipulation of the equations. In other words, according to Maxwell's equations, an electric dipole could be *modelled* as a pair of electric charges on a spring, and the magnetic dipole as an Amperian current-loop (as usual), but the same equations also allow us to model the dipoles differently [7]. For example, they allow us to think of magnetic dipoles as a pair of north and south magnetic monopoles attached to the opposite ends of a spring. Thus, as far as Maxwell's equations are concerned, we do not actually *know* what the dipoles look like. Maxwell's equations allow us to model them, or to imagine them, in different ways. All such models yield precisely the same field-distributions throughout space-time, and therefore, as far as any EM field distributions are concerned, the various models of the dipoles (e.g., Amperian current-loop, Gilbertian dipole, etc.) are equivalent.

If Maxwell's equations fail to pin down the nature of the dipoles, where else can one turn to determine the "correct" model for these dipoles? How about the laws of EM force and torque? Well, there are several laws to choose from; the Lorentz force law is the most famous, of course, but there are alternative laws formulated by Minkowski, Abraham, Chu, and Einstein and Laub [2,8-14]. In general, EM fields act on material media by exerting force and torque on the sources. The two force-density and torque-density expressions that will be the focus of the present paper are due to Lorentz and Einstein-Laub (E-L). In the Lorentz formulation, the EM force and torque densities are given by

$$\boldsymbol{F}_L(\boldsymbol{r},t) = (\rho_{\text{free}} - \boldsymbol{\nabla} \cdot \boldsymbol{P})\boldsymbol{E}(\boldsymbol{r},t) + (\boldsymbol{J}_{\text{free}} + \partial_t \boldsymbol{P} + \mu_0^{-1}\boldsymbol{\nabla} \times \boldsymbol{M}) \times \boldsymbol{B}(\boldsymbol{r},t), \tag{5}$$

$$\boldsymbol{T}_L = \boldsymbol{r} \times \boldsymbol{F}_L. \tag{6}$$

The corresponding expressions in the Einstein-Laub formulation [11] are

$$\boldsymbol{F}_{EL}(\boldsymbol{r},t) = \rho_{\text{free}}\boldsymbol{E} + \boldsymbol{J}_{\text{free}} \times \mu_0 \boldsymbol{H} + (\boldsymbol{P} \cdot \boldsymbol{\nabla})\boldsymbol{E} + \partial_t \boldsymbol{P} \times \mu_0 \boldsymbol{H} + (\boldsymbol{M} \cdot \boldsymbol{\nabla})\boldsymbol{H} - \partial_t \boldsymbol{M} \times \varepsilon_0 \boldsymbol{E}, \tag{7}$$

$$\boldsymbol{T}_{EL}(\boldsymbol{r},t) = \boldsymbol{r} \times \boldsymbol{F}_{EL}(\boldsymbol{r},t) + \boldsymbol{P} \times \boldsymbol{E} + \boldsymbol{M} \times \boldsymbol{H}. \tag{8}$$

The above formulas can be readily derived from the corresponding stress-tensors of classical electrodynamics [7,12], which additionally reveal that the EM momentum density in the Lorentz formulation is $\boldsymbol{p}_L^{(EM)} = \varepsilon_0 \boldsymbol{E} \times \boldsymbol{B}$ (Livens momentum), while that in the E-L formulation is $\boldsymbol{p}_{EL}^{(EM)} = \boldsymbol{E} \times \boldsymbol{H}/c^2$ (Abraham momentum); here $c = (\mu_0 \varepsilon_0)^{-\frac{1}{2}}$ is the speed of light in vacuum.

How does one verify the veracity of a force law? One cannot resort to the familiar models (i.e., pair-of-electric-charges-on-a-spring for $\boldsymbol{p}$, and the Amperian current-loop for $\boldsymbol{m}$) and argue backwards, because, as mentioned earlier, *all* such models comply with Maxwell's equations. Are there experiments that show, for instance, that the predictions of the Lorentz theory are in agreement with experimental results, while those of Einstein and Laub are not? It can be readily shown [7] that the *total* force and *total* torque on any isolated object turn out to be precisely the same in the two formulations (provided that hidden entities are properly discounted in the Lorentz formulation). Nevertheless, there exist experiments that can distinguish between the various theories based on the *distribution* of EM force inside deformable media. In these experiments what is important is *not* total force and total torque, but rather the *distribution* of EM force and torque [15-19]. This is one place to which we might turn our attention in order to compare the various theories against each other. The limited experimental evidence in this direction seems to favor the Einstein-Laub formulation [19].

Another class of relevant experiments involves the hyperfine energy splitting in hydrogen and hydrogen-like atoms [20-22], and also the scattering of neutrons from magnetized matter [22]. In particular, Jackson's paper [22] addresses the question of how to interpret the origin of the *intrinsic* magnetic moments associated with various constituent particles of ordinary atoms as well as those of positronium and muonium. He discusses the intrinsic magnetic moments of the electron, the proton, the positron, the muon, and the neutron, and carries out two kinds of calculations: i) the hyperfine interaction energy between the magnetic moment of an electron and that of a proton (or a positron, or a muon) for hydrogen and hydrogen-like atoms in their $s$-states; ii) the scattering amplitude of neutrons colliding with electrons in ferromagnetic materials.

The basic question posed by Jackson is: Can the intrinsic magnetic moment (associated with spin) be modeled as arising from a circulating electrical current or from a separated north-south pair of magnetic monopoles? The question boils down to whether, in the region where the Schrödinger wave-functions of two particles overlap, the interaction energy $\mathcal{E}$ between the magnetic dipole moment $\boldsymbol{m}$ of one particle and the *internal* magnetic field ($\boldsymbol{B}$ or $\boldsymbol{H}$) of the other particle is given by $\mathcal{E} = -\mu_0^{-1} \boldsymbol{m} \cdot \boldsymbol{B}$ or $\mathcal{E} = -\boldsymbol{m} \cdot \boldsymbol{H}$. (Note that, with $\boldsymbol{B}$ defined as $\mu_0 \boldsymbol{H} + \boldsymbol{M}$, the magnetic moment of a loop of current $I$, area $A$, and surface-normal $\hat{\boldsymbol{n}}$, will be $\boldsymbol{m} = \mu_0 I A \hat{\boldsymbol{n}}$.) The same question may be restated as follows: Is there, in addition to the field-matter interaction energy $\mathcal{E} = -\boldsymbol{m} \cdot \boldsymbol{H}$, a particle-particle interaction energy proportional to $-\boldsymbol{m}_1 \cdot \boldsymbol{m}_2$, when two particles having magnetic moments $\boldsymbol{m}_1$ and $\boldsymbol{m}_2$ are superimposed?

Jackson rightly points out that, for atomic $s$-states, there is no net orbital angular momentum and that the probability-density function of the electron in an $s$-state is fairly confined to the vicinity of the nucleus; in other words, the electron spends a fraction of its time at the location of the nucleus. The hyperfine splitting, therefore, must come from the interaction between the magnetic moment of the electron and the *internal* magnetic field of the nucleus. (The interaction energy between the magnetic moment of the $s$-state electron and the *external* field of the nucleus averages out to zero.) Now, it is well known that the internal fields of a uniformly-magnetized spherical particle of magnetization $\boldsymbol{M}$ are $\boldsymbol{H} = -\boldsymbol{M}/(3\mu_0)$ and $\boldsymbol{B} = 2\boldsymbol{M}/3$ [4,5]. Denoting the magnetic moment of the electron by $\boldsymbol{m}_e$ and that of the nucleus by $\boldsymbol{m}_n$, the hyperfine splitting energy of the *s*-state electron is thus seen to be proportional to ⅔$\bm{m}_e \cdot \bm{m}_n$ for the Amperian model, and to $-⅓\bm{m}_e \cdot \bm{m}_n$ for the Gilbertian model of magnetization. Not only are the energies obtained for the two models opposite in sign, but also they differ in magnitude by a factor of two. The spectroscopy results for hydrogen and hydrogen-like atoms analyzed by Jackson and others [20-22] clearly favor the Amperian (against the Gilbertian) model for intrinsic magnetic dipoles.

In the remainder of this paper we discuss the similarities and differences between the Lorentz and Einstein-Laub formulations of classical electrodynamics. Both formalisms have their merits, but could lead to confusion and misapplication if not fully and properly understood.

**2. Treating *P* and *M* as independent sources, on a par with $\rho_{\text{free}}$ and $\bm{J}_{\text{free}}$.** It has been argued that if one takes, as the starting point, Maxwell's macroscopic equations for the $\bm{E}, \bm{D}, \bm{H}, \bm{B}$ fields, one would need the constitutive relations for $\bm{D}$ and $\bm{H}$ in terms of $\bm{E}$ and $\bm{B}$ to be able to solve the equations for given $\rho_{\text{free}}$ and $\bm{J}_{\text{free}}$ (and boundary conditions). However, if $\bm{P}$ and $\bm{M}$ were taken as independent sources, along with $\rho_{\text{free}}$ and $\bm{J}_{\text{free}}$, then the ensuing theory would diverge from the Maxwell-Lorentz electrodynamics, which considers only charge and current as the ultimate sources of the EM field. Such an electrodynamics presumably would have four fundamental sources, and the $(\bm{E}, \bm{B})$ fields will not be its only fundamental fields. Operationally and/or experimentally, however, since one has control only over charges and currents, setting up a given $\bm{P}$ and/or $\bm{M}$ independently of the former sources will be problematic.

Contrary to the above assertions, taking $\bm{P}$ and $\bm{M}$ as independent sources, similar to $\rho_{\text{free}}$ and $\bm{J}_{\text{free}}$, does not make any difference as far as Maxwell's equations are concerned. All constitutive relations are still applicable, and whether $\bm{P}$ and $\bm{M}$ behave independently of the external fields, or depend linearly or nonlinearly on the external fields, etc., makes no difference in the treatment of Maxwell's equations. One puts $\bm{P}$ and $\bm{M}$ into the equations, then replaces $\bm{P}$ and $\bm{M}$ using the constitutive relations (however simple or complicated these relations may be), and proceeds to solve the equations. If one *chooses* to call $-\nabla \cdot \bm{P}$ the bound-charge-density $\rho_{\text{bound}}$, and $\partial_t \bm{P} + \mu_0^{-1} \nabla \times \bm{M}$ the bound-current-density, $\bm{J}_{\text{bound}}$, then one will have exactly the same equations (in terms of free and bound charges and currents) as everyone else will have. If, on the other hand, one chooses to rearrange Maxwell's equations differently so that magnetic bound-charge, $-\nabla \cdot \bm{M}$, and magnetic bound-current, $\partial_t \bm{M}$, appear in the equations, then the equations will look superficially different from the standard *microscopic* equations. However, this will have no effect on the solutions, because one is still solving the same equations and, in the end, one will get the same $\bm{E}, \bm{D}, \bm{H}, \bm{B}$ fields that everyone else will get out of Maxwell's equations. In other words, whatever one *thinks* of the electric and magnetic dipoles (i.e., as Amperian current loops, or as Gilbertian pairs of north-south poles, or as magnetic monopoles going round and round in circles, etc.), it will *not* matter, because Maxwell's equations remain the same and, therefore, their ultimate solutions will turn out to be exactly the same.

Where the treatment of $\bm{P}$ and $\bm{M}$ as independent sources does make a difference is in the evaluation of EM force and torque. Again, what matters here is the end result of calculations (and its agreement with experiments). It does not matter at all how one may *feel* about the nature of the dipoles; what is important is the equations that one will use and the numbers that one will get out of these equations—numbers for force, torque, momentum, angular momentum, etc. Now, if one uses the Lorentz law of force/torque, one will get a certain set of numbers which differ from those that one gets out of the E-L equations. Of course, one will have to discount the effects of hidden momentum [23-29], i.e., subtract the hidden contributions to force and torque, before the Lorentzian results can be used in conjunction with Newton's law ($\bm{F} = m\bm{a}$, or its relativistic version). In contrast, the E-L values do not need to go through this additional step of discounting the hidden entities. Once hidden entities are removed, the Lorentz formalism will give *precisely* the same *total* force and *total* torque on any isolated object as one would get from the E-L equations [7,30-35]. Therefore, experiments on isolated solid objects cannot tell the difference between the two formalisms. Experiments *can* tell the difference between the two if the *distribution* of force/torque inside deformable media is examined [15-19]. That is why the effect of EM radiation on deformable media is an important area for experimental investigation.

**3. The case of a uniformly-polarized cylinder in an external electric field**. A comparison between the Lorentz and E-L force densities in the simple case of a permanently uniformly polarized cylinder in a uniform electric field is instructive. Bound charges appear on the top and bottom facets of the cylinder. If the external uniform field $\boldsymbol{E}$ is parallel to the cylinder's axis, then this surely will subject the cylinder to a compressive or a tensile stress, depending on the direction of $\boldsymbol{E}$, whereas the E-L force-density vanishes everywhere inside and on the surface of the cylinder. Does this mean that, according to the E-L formulation, the cylinder will not be subject to any stresses? The answer is no; the E-L formula only suggests that individual dipoles, each considered as a single entity, will not be subject to a *net* force from the uniform, externally applied $E$-field. Of course, there will be compressive or tensile stresses built up *within* each dipole, but what happens inside a dipole is not the subject of the E-L formulation.

Since the E-L theory treats dipoles as indivisible entities, one cannot expect the inner workings of a dipole to emerge from this formulation. One could, of course, use a model, such as the ball-and-stick model, for an electric dipole. One will then have specified the dipole in terms of its constituents—which, in this case, is a pair of equal and opposite electric charges separated by a stick or a spring. The first two terms in the E-L force-density of Eq.(7), $\rho \boldsymbol{E} + \boldsymbol{J} \times \mu_0 \boldsymbol{H}$, will then allow one to examine the internal stresses in the presence of external fields. Clearly, a model is needed to reveal the internal workings of the dipole.

Note also that the same uniformly-polarized cylinder in the presence of a *non-uniform* externally applied $E$-field will behave differently in the two formulations. In the Lorentz approach, the forces are localized at the top and bottom of the cylinder, where $\boldsymbol{\nabla} \cdot \boldsymbol{P} \neq \boldsymbol{0}$, whereas in the E-L formalism the EM force is distributed throughout the volume.

One may expand the interpretation of the Lorentz force by arguing that forces are transmitted as mechanical stresses through the body of the cylinder [36]. Denoting the surface normal by $\boldsymbol{n}$, there will be stresses $(\boldsymbol{P} \cdot \boldsymbol{n})\boldsymbol{E}$ on any surface drawn within the cylinder, even when $\boldsymbol{\nabla} \cdot \boldsymbol{P} = 0$ inside the cylinder. Here $\boldsymbol{P} \cdot \boldsymbol{n} = \sigma_s$ is the bound charge-density on any part of the chosen (internal) surface. Using the tensor identity $\overleftrightarrow{\boldsymbol{\nabla}} \cdot (\boldsymbol{PE}) = (\boldsymbol{P} \cdot \boldsymbol{\nabla})\boldsymbol{E} + (\boldsymbol{\nabla} \cdot \boldsymbol{P})\boldsymbol{E}$, and invoking Gauss's theorem (i.e., $\int_V \boldsymbol{\nabla} \cdot \boldsymbol{A}\, dv = \oint_S \boldsymbol{A} \cdot d\boldsymbol{s}$), we will have

$$F_i = \int_{Volume}(\boldsymbol{P} \cdot \boldsymbol{\nabla})E_i\, dv = -\int (\boldsymbol{\nabla} \cdot \boldsymbol{P})E_i\, dv + \int \boldsymbol{\nabla} \cdot (E_i \boldsymbol{P})dv$$
$$= -\int_{Volume}(\boldsymbol{\nabla} \cdot \boldsymbol{P})E_i\, dv + \oint_{Surface}(\boldsymbol{P} \cdot \boldsymbol{n})E_i\, ds, \quad (i = x, y, z). \qquad (9)$$

In words, it is the integral of $(-\boldsymbol{\nabla} \cdot \boldsymbol{P})\boldsymbol{E}$ over any volume inside the cylinder, augmented by the integral of $(\boldsymbol{P} \cdot \boldsymbol{n})\boldsymbol{E}$ over the volume's closed surface, that should be regarded as the true Lorentz force experienced by the dipoles within the specified volume [36]. This interpretation of the Lorentz force brings it into complete accord with the E-L force, which appears on the left-hand side of Eq.(9). We mention in passing that this alternative interpretation of the Lorentz force on electric dipoles calls for a corresponding change in the torque-density formula, as the E-L torque-density of Eq.(8) is seen to have an additional term in the form of $\boldsymbol{P} \times \boldsymbol{E}$.

**4. Gilbertian dipoles have the same contact-term as Amperian dipoles**. It has been suggested that the magnetic fields of Amperian and Gilbertian dipoles differ substantially within the dipole's interior. In the idealization of a point-dipole, this is expressed by different contact terms of the field, which acquire the form of $\delta$-functions at the location of the dipole. Perhaps some misunderstanding of the nature of Gilbert dipoles causes this confusion. To be absolutely clear, let us state once and for all that the internal fields of Amperian and Gilbertian dipoles are precisely the same, as are indeed their external magnetic fields.

Imagine a uniformly-magnetized sphere having magnetization-density $\boldsymbol{M}$ along the $z$-axis. In the Lorentz model, the $B$-field inside the sphere, produced by the Amperian current circulating around the exterior surface of the sphere, is found to be $2\boldsymbol{M}/3$. In the E-L model, the $H$-field inside the sphere, produced by the opposite magnetic charge-densities residing on the upper and lower hemispherical surfaces, is found to be $-\boldsymbol{M}/(3\mu_0)$. Since $\boldsymbol{B} = \mu_0 \boldsymbol{H} + \boldsymbol{M}$, it is clear that $\boldsymbol{B} = 2\boldsymbol{M}/3$ in the E-L case and that, therefore, the two models are in perfect agreement. In the limit when the radius of the sphere shrinks to zero, both models predict the same contact term in the form of a $\delta$-function. These contact terms cannot possibly be different, because the Lorentz dipoles and the Einstein-Laub dipoles obey *the same* equations of Maxwell.

**5. Can macroscopic stress-strain relations distinguish Amperian from Gilbertian dipoles?** One might suspect that the strains in polarized or magnetized bodies due to their internal fields themselves should shed light on the distributions of the electric and magnetic forces. When a magnetized bar is modeled as circulating surface currents, there should be a pressure on its surface similar to that on a current-carrying solenoid. But when it is modeled as magnetic poles/charges on its ends, there should be pressure on the ends similar to that in a charged parallel-plate capacitor. Here perhaps the well-studied phenomena of magnetostriction and electrostriction are relevant. A uniformly-polarized bar is strained in a manner compatible with forces on surface charges on its ends, exerted by the internal $E$-field of the bar. And when a rod is uniformly magnetized, it is strained in a manner compatible with forces on circular currents on its surface, exerted by the $B$-field within the rod. If there are Amperian currents on the surface of a magnetized cylindrical rod, then the action of the internal $B$-field on these currents will subject the rod's surface to a tensile stress that, given a typical elastic modulus of the rod's material, should increase its radius and decrease its length. And if there are Gilbertian poles on the ends of the rod, surely the action of the internal $B$-field will subject the ends to a compressive stress that will shorten the rod's length and increase its average radius accordingly. The overall deformation, while qualitatively similar, may yet turn out to be different from that expected on the Amperian model. But if the elastic moduli depend significantly on the magnetization (or polarization), then the analysis will not be straightforward, nor will it get any easier when external fields act on magnetized (or polarized) bodies [36].

If we take a cylinder of ferroelectric material and, starting in a depolarized state, bring it to a saturated state along the axis of the cylinder, upon removing the external $E$-field and in the absence of interactions between the dipoles and the crystal lattice, we would expect the bar's internal $E$-field to try to shorten the cylinder (i.e., the "capacitor plates" would attract each other). In similar fashion, if we take a ferromagnetic cylinder and, starting in a demagnetized state, bring it to a state of saturation along the cylinder axis, then, upon removing the external $B$- field and in the absence of interactions between the dipoles and the crystal lattice, we would expect the internal *B*-field to once again try to shorten the cylinder—this time, with the internal *B*-field acting on the solenoidal current around the cylinder. The latter effect cannot, therefore, distinguish between (i) the pressure of the internal *B*-field on the Amperian current, and (ii) the pull of the north-south poles of the bar magnet on each other.

Electrostriction and magnetostriction, of course, are complex solid-state phenomena involving ferroelectric and ferromagnetic domains and their domain walls, as well as atomic-level interactions between dipoles and the crystalline lattice. In the case of magnetostriction, for example, although the first observations (by James Prescott Joule in 1842) showed a negative coefficient of magnetostriction (i.e., shortening of a rod along the direction of an applied *B*-field), subsequent investigations showed that many ferro- and ferri-magnetic materials exhibit a positive coefficient. Longitudinal magnetostriction, therefore, can be positive or negative, depending on the nature and composition of the magnetic material and its microstructure [37]. In the case of electrostriction, the longitudinal coefficient is generally positive, although the phenomenon has a strong dependence on the applied *E*-field as well as on the intrinsic properties of the material. All in all, it appears that macroscopic stress-strain relations cannot be used to argue in favor of one or the other model of magnetic dipoles.

**6. Imposing the perceived nature of magnetic dipoles on the electromagnetic force law**. In their 1908 paper [11], Einstein and Laub explain that they have chosen the magnetization-dependent part of the EM force-density to be entirely analogous to the polarization-dependent part. In other words, since the EM force on an electric dipole $\boldsymbol{p}$ at rest is

$$\boldsymbol{F}_p = (\boldsymbol{p} \cdot \boldsymbol{\nabla})\boldsymbol{E} + (\partial_t \boldsymbol{p}) \times \mu_0 \boldsymbol{H}, \tag{10}$$

they suggest that the force on a magnetic-dipole $\boldsymbol{m}$ at rest must be

$$\boldsymbol{F}_m = (\boldsymbol{m} \cdot \boldsymbol{\nabla})\boldsymbol{H} - (\partial_t \boldsymbol{m}) \times \varepsilon_0 \boldsymbol{E}. \tag{11}$$

However, since the magnetic dipoles found in nature are believed to be due to electric-current loops, the force on them, in accordance with the Lorentz force law, is expected to be

$$\boldsymbol{F}_m = \mu_0^{-1} \boldsymbol{\nabla}(\boldsymbol{m} \cdot \boldsymbol{B}) = \mu_0^{-1}(\boldsymbol{m} \cdot \boldsymbol{\nabla})\boldsymbol{B} + \boldsymbol{m} \times (\mu_0^{-1} \boldsymbol{\nabla} \times \boldsymbol{M} + \varepsilon_0 \partial_t \boldsymbol{E}). \tag{12}$$

The simplifying assumptions in arriving at Eq.(12) are that, at the location of the dipole $\boldsymbol{m}$, free charge-density $\rho_{\text{free}}$, free current-density $\boldsymbol{J}_{\text{free}}$, and polarization $\boldsymbol{P}$ are absent, leaving only the local *E*-field, $\boldsymbol{E}(\boldsymbol{r},t)$, and the local *B*-field, $\boldsymbol{B}(\boldsymbol{r},t) = \mu_0 \boldsymbol{H} + \boldsymbol{M}$, to exert an EM force $\boldsymbol{F}_m$ on the static dipole $\boldsymbol{m}$. It is tempting to argue that, for the simple reason that Eq.(12) differs from Eq.(11), the analogy with electric dipoles that Einstein and Laub used cannot be correct.

However, one should ask whether the magnetic dipoles found in nature are *in every respect* analogous to electric current loops. For orbital magnetic moments, the current-loop analogy is apt but imperfect, given the quantum-mechanical nature and the stable character of the loop, which are surely beyond the bounds of classical physics. As for the intrinsic magnetic moments of elementary particles arising from their intrinsic spin—an undeniably quantum phenomenon—the analogy may be even more problematic, despite experimental evidence that apparently rules out the Gilbertian magnetic-charge model by suggesting that these moments can be accurately modeled by circulating electrical currents [20-22].

Without discounting the strong evidence in favor of the current-loop model, one should resist the temptation to leap to the conclusion that the Amperian model is applicable in every respect and under all circumstances. For instance, one may inquire: what is the diameter of the current loop associated with the magnetic dipole of a free electron? How fast does the negative charge of the electron rotate around its spin axis? Is this current due entirely to negative charge, or could there perhaps be both positive and negative charges (albeit in different amounts) inside the electron, rotating in opposite directions, at different radii, and at different speeds? In other words, how certain can one be of the underlying structure of the electron's magnetic moment?

A similar objection may be raised to our presumed knowledge of the electric dipole moment $\boldsymbol{p}$. An electron with a magnetic moment $\boldsymbol{m}$ traveling with velocity $\boldsymbol{V}$ will have a relativistically-induced electric dipole moment $\boldsymbol{p} = \varepsilon_0 \boldsymbol{V} \times \boldsymbol{m}$. What does this electric dipole look like? Has the interior of an electron been examined to reveal the magnitude and distance between the pair of charges that are presumably separated from each other inside the electron to give rise to this $\boldsymbol{p}$?

The point is that, while under certain circumstances, we may know something about the nature of electric and magnetic dipoles, we cannot always be certain that what has traditionally been *assumed* about these dipoles is valid under every circumstance. In particular, extending the analogy between a magnetic dipole and an ordinary current-loop to cover the force and torque exerted by EM fields on the dipole, while plausible, may not necessarily be correct. We must, therefore, examine the subtle differences between Eqs.(11) and (12) in some detail. This is the subject of the following section.

**7. Force exerted by local $\boldsymbol{E}$ and $\boldsymbol{B}$ fields on a stationary magnetic point-dipole $\boldsymbol{m}$**. The main difference between the Amperian and Gilbertian models of magnetic dipoles shows up when a magnetic point-dipole $\boldsymbol{m}$ is acted upon by local $\boldsymbol{E}$ and $\boldsymbol{B}$ fields, with $\boldsymbol{B}(\boldsymbol{r},t) = \mu_0 \boldsymbol{H}(\boldsymbol{r},t) + \boldsymbol{M}(\boldsymbol{r},t)$ incorporating the magnetization $\boldsymbol{M}(\boldsymbol{r},t)$ of a host medium within which the point-dipole $\boldsymbol{m}$ resides. For the sake of simplicity, let us assume that, at the location of the dipole, $\rho_{\text{free}} = 0$, $\boldsymbol{J}_{\text{free}} = 0$, and $\boldsymbol{P} = 0$. The dipole's magnetization, $\boldsymbol{M}_0(\boldsymbol{r},t) = \boldsymbol{m}\delta(x)\delta(y)\delta(z)$, is associated with its bound current-density $\boldsymbol{J}_{\text{bound}} = \mu_0^{-1}\boldsymbol{\nabla} \times \boldsymbol{M}_0$, which leads to the Lorentz force $\int \boldsymbol{J}_{\text{bound}} \times \boldsymbol{B} dv$. Using the fact that $\boldsymbol{\nabla} \cdot \boldsymbol{B} = 0$, the end result may be simplified, and the Lorentz force on the dipole written as $\mu_0^{-1}\boldsymbol{\nabla}(\boldsymbol{m} \cdot \boldsymbol{B})$, with the caveat that $\boldsymbol{m}$ is treated as a constant, while the gradient operator acts on $\boldsymbol{B}(\boldsymbol{r},t)$. Upon subtracting the hidden momentum contribution [23-29], the Lorentz force (understood as mass × acceleration) exerted on the Amperian dipole becomes

$$\begin{aligned}\boldsymbol{F}_m &= \mu_0^{-1}\boldsymbol{\nabla}(\boldsymbol{m} \cdot \boldsymbol{B}) - \partial_t(\boldsymbol{m} \times \varepsilon_0 \boldsymbol{E}) \\ &= \mu_0^{-1}[(\boldsymbol{m} \cdot \boldsymbol{\nabla})\boldsymbol{B} + \boldsymbol{m} \times (\boldsymbol{\nabla} \times \boldsymbol{B})] - \partial_t(\boldsymbol{m} \times \varepsilon_0 \boldsymbol{E}) \\ &= (\boldsymbol{m} \cdot \boldsymbol{\nabla})\boldsymbol{H} + \mu_0^{-1}(\boldsymbol{m} \cdot \boldsymbol{\nabla})\boldsymbol{M} + \boldsymbol{m} \times (\varepsilon_0 \partial_t \boldsymbol{E} + \mu_0^{-1}\boldsymbol{\nabla} \times \boldsymbol{M}) - \partial_t(\boldsymbol{m} \times \varepsilon_0 \boldsymbol{E}) \\ &= (\boldsymbol{m} \cdot \boldsymbol{\nabla})\boldsymbol{H} - (\partial_t \boldsymbol{m}) \times \varepsilon_0 \boldsymbol{E} + \mu_0^{-1}[(\boldsymbol{m} \cdot \boldsymbol{\nabla})\boldsymbol{M} + \boldsymbol{m} \times (\boldsymbol{\nabla} \times \boldsymbol{M})]. \end{aligned} \qquad (13)$$

The difference between the above equation, which gives the Lorentz force on a stationary magnetic point-dipole $\boldsymbol{m}$ (minus the contribution of hidden momentum), and the corresponding E-L force on a Gilbertian dipole is the last term on the right-hand-side of Eq.(13), namely, $\mu_0^{-1}[(\boldsymbol{m} \cdot \boldsymbol{\nabla})\boldsymbol{M} + \boldsymbol{m} \times (\boldsymbol{\nabla} \times \boldsymbol{M})]$. Evidently, the complete force expression in Eq.(13) is needed to explain the hyperfine energy-splitting in hydrogen and hydrogen-like atoms mentioned earlier, as well as experiments on the scattering of polarized neutrons from ferromagnetic materials [22].

In such problems one must be careful with interpretation, because *two* sources of magnetization, namely, the magnetic point-dipole $\boldsymbol{m}$ (associated in the neutron scattering experiments with the neutron), and the magnetization $\boldsymbol{M}$ of the host material, are co-located in space. The E-L expression of force-density pertains to the *total* force exerted on *both* sources of magnetization, and, as pointed out elsewhere [7], the E-L expression may be parsed in different ways in order to allocate different shares to the various participants in the process.

Suppose a magnetic point-dipole $\boldsymbol{m}_1$ is co-located in space with another magnetic point-dipole $\boldsymbol{m}_2$. If the aforementioned additional force-density $\mu_0^{-1}[(\boldsymbol{M}_1 \cdot \boldsymbol{\nabla})\boldsymbol{M}_2 + \boldsymbol{M}_1 \times (\boldsymbol{\nabla} \times \boldsymbol{M}_2)]$ acts on the first dipole, there will be a corresponding force-density $\mu_0^{-1}[(\boldsymbol{M}_2 \cdot \boldsymbol{\nabla})\boldsymbol{M}_1 + \boldsymbol{M}_2 \times (\boldsymbol{\nabla} \times \boldsymbol{M}_1)]$ acting on the second dipole. The sum of these force-densities will then be $\mu_0^{-1}\boldsymbol{\nabla}(\boldsymbol{M}_1 \cdot \boldsymbol{M}_2)$, which integrates to zero over the (common) volume of the two dipoles. Therefore, as far as the *total force* on the volume of interest is concerned, these additional forces (which are necessary for Jackson's argument in [22]) do *not* alter the Einstein-Laub formula. In the neutron scattering experiments discussed by Jackson in [22], the additional neutron-electron interaction term would be $-\mu_0^{-1}\boldsymbol{m}_1 \cdot \boldsymbol{M}_2$; see Jackson's Eqs.(23) and (24).

Note that it is *not* the need for the term $\mu_0^{-1}[(\boldsymbol{m} \cdot \boldsymbol{\nabla})\boldsymbol{M} + \boldsymbol{m} \times (\boldsymbol{\nabla} \times \boldsymbol{M})]$ in Eq.(13) that is in dispute here. What we have is a particle with dipole moment $\boldsymbol{m}$ (e.g., a neutron) passing through an external medium, say, an electron, which itself has magnetization $\boldsymbol{M}$. When the wave functions of the two particles overlap, the force $\mu_0^{-1}[(\boldsymbol{m} \cdot \boldsymbol{\nabla})\boldsymbol{M} + \boldsymbol{m} \times (\boldsymbol{\nabla} \times \boldsymbol{M})]$ acting on the first particle will be equal and opposite to the force acting on the second (due to the magnetization of the first particle). The *total force* acting on the pair of particles—one presumably residing inside the other—is therefore zero. Thus the E-L formula, which only pertains to the *total force* on the *combination* of the two particles (i.e., neutron inside a magnetized host) remains operative. However, calculation of scattering amplitudes requires taking into account the particle-particle interaction energy $-\mu_0^{-1}\boldsymbol{m} \cdot \boldsymbol{M}$.

Of course, when a particle moves through a medium, the *self-force* on its charge and also on its dipole moment(s) become relevant as well. In the preceding example, the EM force exerted by the *H*-field of the neutron acting on itself can no longer be ignored. So long as the neutron is unaccelerated (i.e., no acceleration in its linear motion, nor any temporal change in its dipole moment's orientation), the total self-force will be zero. This is true simply because unaccelerated objects should not exert a force on themselves—but it can also be proven rigorously. The self-field and the self-force are present in both the Lorentz and E-L formulations; in both cases the net self-force acting on an unaccelerated particle turns out to be zero. When the particle is accelerated, it will radiate and, consequently, its self-force will no longer vanish; this is known as radiation resistance. Not only in the specific case of the scattering of neutrons from a magnetic material, but also in general, whether one uses the Lorentz or the E-L formalism, the $\boldsymbol{E}, \boldsymbol{D}, \boldsymbol{H}, \boldsymbol{B}$ fields used in the calculations of force and/or torque must be the *total* fields, that is, the external fields plus the fields produced by the moving particle itself.

Finally, it is perhaps worth mentioning at this point that the interaction between a moving neutron and its host medium is, in general, a complicated problem involving collision cross-sections, spin-spin and spin-orbit couplings, the action of the lattice *E*-field on the (relativistically-induced) electric dipole accompanying a moving magnetic dipole, etc., none of which may be easily expressible in terms of EM fields and sources. The situation is perhaps similar to the case of extraordinary (or anomalous) Hall effect, where a charged particle (electron) moves through a ferromagnetic host. It is well known that the sign and the magnitude of the extraordinary Hall voltage cannot be expressed in terms of the Lorentz force acting on the moving charged particles alone—in other words, the interactions of the electrons with the lattice are also important factors in determining the observed Hall voltage [38].

**8. Field-matter and matter-matter interactions**. The word "force" used in both the Lorentz and E-L formulations of classical electrodynamics could be misleading. For instance, the Lorentz force is *not* necessarily the same force as understood in the context of Newtonian mechanics (or the relativistic extension of it), unless the contribution of hidden momentum is removed from it. The EM force in all the various formulations of the classical theory (Lorentz, Einstein-Laub, Chu, Minkowski, Abraham, etc.) is simply the rate-of-exchange of linear momentum between the fields and the material media. While the EM momentum-density in the Lorentz formulation is the so-called Livens momentum-density, $\varepsilon_0 \boldsymbol{E} \times \boldsymbol{B}$, the same entity is given by $\boldsymbol{D} \times \boldsymbol{B}$ in Minkowski's formulation, and by $\boldsymbol{E} \times \boldsymbol{H}/c^2$ (Abraham's momentum-density) in the Einstein-Laub, Chu, and Abraham formulations. In each case, the corresponding force equation provides a formula for calculating the local rate-of-exchange of the EM momentum with the material media—the material media being the seats of $\rho_{\text{free}}, \boldsymbol{J}_{\text{free}}, \boldsymbol{P}$ and $\boldsymbol{M}$.

Now, if the material media remain more or less motionless (as in the case of reflection of light from a massive mirror), we speak of radiation pressure, and the aforementioned formulas can be used to confirm the conservation of momentum. (The mirror does not move perceptibly, but, because of its large mass, it picks up mechanical momentum at the expense of the EM momentum of the radiation field.) If, however, the material medium happens to be an isolated solid object whose interaction with EM fields would affect its state of motion, then it must be determined whether the acquired momentum is hidden or not. The part that is not hidden contributes to the motion, although, even in this case, the equations of motion are not simple, because the EM energy stored within the object could modify its rest-mass [39].

If the material medium happens to be deformable, then the *distribution* of EM force-density becomes relevant, in which case the EM theory alone is not sufficient to predict the behavior of the material medium—equations of elasticity or hydrodynamics must now be called upon to predict the evolution of the system. Clearly, other forces now come into play that have nothing to do with the *original* EM forces, but are related to the action of each part of the medium on its neighboring parts. Whatever the nature of these forces may be, their actions on the various parts of the material medium will be equal and opposite, so that no additional momentum will be imparted to the medium as a whole in response to these internal forces. Each part, however, may acquire momentum above and beyond what was originally imparted to it by the EM fields.

Finally, if the interactions involve the passage of one particle through another (e.g., a charged particle going through a magnetic dipole, or one magnetic dipole passing through another magnetic dipole), we will have particle-particle interaction in addition to field-matter interaction. The E-L theory describes the field-matter interaction part, and the particle-particle interactions will have to be treated separately and in their own rights. As in the aforementioned case of deformable media, the particles could exchange energy and momentum with each other, *without* affecting the overall balance of energy and momentum between the fields and the material media, the latter being governed by the electromagnetic laws of force, torque, energy, and momentum.

When two hard spheres (both uncharged and non-magnetic) collide, we do not, for obvious reasons, describe their scattering in electromagnetic terms. Similarly, the overlap energy of two magnetic point-dipoles (which is proportional to $-\boldsymbol{m}_1 \cdot \boldsymbol{m}_2$) is what is needed to describe the hard-sphere-like scattering of one dipole from the other. In Jackson's analysis of the scattering of neutrons from a magnetized medium [22], this contact energy $(-\boldsymbol{m}_1 \cdot \boldsymbol{m}_2)$ must be added to the interaction energy between the *H*-field and the particle, the latter energy being readily accounted for by the E-L formula. It appears that a combination of field-particle interaction (Einstein-Laub) and particle-particle scattering is the proper way to handle situations involving the passage of a magnetic particle (or that of a charged particle) through a magnetic medium.

The E-L formula describes the time-rate-of-exchange of linear momentum between EM fields and material media; it is silent on the question of momentum exchange between one part of a material medium and another (e.g., in the case of deformable media), or the momentum exchange between one material particle superimposed onto another (e.g., hard-sphere-like scattering of neutrons upon colliding with a magnetic material's electrons).

This is not unlike the situation with the Lorentz law. Strictly speaking, the Lorentz force-density is given by Eq.(5), where $\boldsymbol{E}$ and $\boldsymbol{B}$ are the *total* electric and magnetic fields at the point $(\boldsymbol{r}, t)$ in spacetime; similarly, $\rho_{\text{free}}$, $\boldsymbol{J}_{\text{free}}$, $\boldsymbol{P}$ and $\boldsymbol{M}$ are the total charge, current, polarization, and magnetization densities at the given point. In the case of deformable media, this Lorentz law is also silent on the question of momentum imparted by one part of the material medium to its adjacent parts, e.g., due to elastic forces, hydrodynamic pressure, tensile stresses, friction forces, quantum-mechanical exchange interactions, etc.

Anything else that one might read into the *pure* Lorentz law of Eq.(5) (e.g., contributions from hidden momentum) requires additional arguments/postulates. For instance, when two or more material particles reside at the same space-time location $(\boldsymbol{r}, t)$, how should one go about parsing the Lorentz force expression of Eq.(5) in order to determine not only the EM force exerted by $\boldsymbol{E}$ and $\boldsymbol{B}$ on *individual* particles, but also any mutual attractive/repulsive forces that one particle might exert on the other(s)? The unspoken assumption here seems to have been that every particle has its own share of $\rho_{\text{free}}$, $\boldsymbol{J}_{\text{free}}$, $\boldsymbol{P}$ and $\boldsymbol{M}$, and that there are no additional forces on any given particle above and beyond what the plain interpretation of the aforementioned form of the Lorentz law suggests. Without quarreling with this assumption, it must be pointed out that the experimental evidence for its validity is rare and also limited to very special circumstances, such as those discussed in [22].

It is possible, of course, to separate field-matter interaction from particle-particle interaction, and this is precisely what is done in [22]; Jackson's treatment involves writing the interaction energy between two particles as the sum of two terms, one for the action of the $\boldsymbol{H}$ field on $\boldsymbol{m}$, the other for the action of the magnetization $\boldsymbol{M}$ on $\boldsymbol{m}$. For two magnetic point-dipoles $\boldsymbol{m}_1$ and $\boldsymbol{m}_2$, the hard-sphere-like particle-particle interaction energy is seen to be proportional to $-\boldsymbol{m}_1 \cdot \boldsymbol{m}_2$. This energy is directly related to the force term $\mu_0^{-1}[(\boldsymbol{m} \cdot \boldsymbol{\nabla})\boldsymbol{M} + \boldsymbol{m} \times (\boldsymbol{\nabla} \times \boldsymbol{M})]$ in Eq.(13). Stated differently, the hard-sphere-like interaction energy of the particles being written as proportional to $-\boldsymbol{m}_1 \cdot \boldsymbol{m}_2$ is a substitute for the force-density term $\mu_0^{-1}[(\boldsymbol{m} \cdot \boldsymbol{\nabla})\boldsymbol{M} + \boldsymbol{m} \times (\boldsymbol{\nabla} \times \boldsymbol{M})]$ in Eq.(13).

In contrast, the E-L force-density equation, being a field-matter interaction equation, is silent on the particle-particle interaction terms. Any scattering process involving one or more material particles must incorporate (in addition to the field-matter energy inherent in the E-L equation) all the requisite energy expressions needed to account for hard-sphere-like scatterings of the participating particles.

Strictly speaking, Jackson's analysis is not a classical treatment; it involves Schrödinger's wave function and certain quantum-mechanical arguments [22]. This, in fact, is how such scattering problems should be treated in general, separating the field-matter interaction from the hard-sphere-like scattering of one particle from another. The E-L formalism is relevant only to the field-matter interaction part. Anytime that a particle possessing electric charge, electric and/or magnetic dipole-moments, etc., collides with another such particle, the corresponding particle-particle interaction energy (e.g., the energy proportional to $-\boldsymbol{m}_1 \cdot \boldsymbol{m}_2$ in the case of two magnetic point-dipoles) must be added in order to determine the resulting scattering amplitude.

In a typical experimental situation, other hard-sphere-like collisions may be involved (e.g., neutron collisions with the atomic nuclei), or the exchange interaction (rooted in Pauli's exclusion principle) may play a role, or the hard-sphere-like collision may be inhibited by the strong repulsion of one charged particle for another, and so on. Under such circumstances, it is perhaps advisable to have separate terms for the field-matter interaction energy on the one hand, and matter-matter interaction energy on the other. The case of slow neutron scattering from a magnetic material's electrons is somewhat special, considering that the neutrons are uncharged and can readily pass through the electrons. In a more complex situation, such as the case of extraordinary (or anomalous) Hall effect, particle-particle interactions are not nearly as straightforward [38]. While a proper treatment of the field-particle and particle-particle scattering (*a la* Jackson) is probably the best way to approach such problems, a straightforward application of the Lorentz force law could be problematic, particularly if it is based on the assumption that the conduction electrons move straight through the magnetic electrons (which are responsible for the ferromagnetism of the host material) and, in the process, bend their paths in response not only to the *H*-field but also in response to the magnetization ***M***. This may be the reason why the extraordinary Hall effect is such a difficult problem, refusing to obey a direct application of the Lorentz force law [38].

**9. Concluding remarks**. Maxwell's *macroscopic* equations, when taken at face value, assume the existence of charge, current, polarization, and magnetization on equal footing. This is useful, because *classical* physics cannot explain the existence of spin and orbital magnetic moments, nor in fact can it explain the existence of atoms and molecules, which are necessary if electric and (orbital) magnetic dipoles are brought into the discussion. Maxwell's equations thus bring ***P*** and ***M*** into existence by fiat, simply by postulating their existence.

Maxwell's equations do *not* assume any specific models for ***P*** and ***M***. For instance, according to these equations, magnetic dipoles may be loops of current (Amperian), or they could consist of pairs of north and south magnetic monopoles (Gilbertian). Similarly, electric dipoles can be pairs of positive and negative charges, or they could be loops of magnetic current produced by circulating magnetic monopoles. All these models are viable according to Maxwell's macroscopic equations; however, the equations do not require one to pick and choose among these models. Whatever happens to be the "correct" internal structure of the dipoles will be fine with Maxwell; all he cares to tell us is the relation between electrical charge and current, polarization, and magnetization on the one hand, and the ***E***, ***D***, ***H***, ***B*** fields on the other.

One does not have to be perplexed as to how to imagine ***P*** and ***M*** as "fundamental" outside of some arrangement of static charges or distribution of currents. One could, for example, contemplate a single electron having a magnetic dipole moment ***m*** moving with velocity ***V*** and acquiring, in accordance with special relativity, an electric point-dipole $\boldsymbol{p} = \varepsilon_0 \boldsymbol{V} \times \boldsymbol{m}$. There is no internal structure here, and it is perhaps impossible to know for sure the distribution of electrical charges and currents inside the electron. Nevertheless, Maxwell's equations, in combination with special relativity and with the E-L force and torque laws, allow for such point-dipoles to exist and to interact with EM fields. This is perhaps as close as one could get to viewing ***p*** and ***m*** as *fundamental* building blocks of classical electrodynamics.

The Lorentz force law applies to electrical charges and currents. In order to apply the law to electric and magnetic dipoles, one must invoke models for ***p*** and ***m***. These turn out to be the pair-of-charges-on-a-stick model for ***p***, and the Amperian current-loop model for ***m***. Often, the end-result of such force calculations is not directly applicable in connection with Newton's second law. That is why hidden momentum must be brought in and its contribution subtracted from the Lorentz force on a magnetic dipole before Newton's law of motion could be applied.

In contrast to the Lorentz approach, the E-L formulation does *not* depend on any models for polarization and magnetization. These sources of the EM field simply exist; they are represented by vector functions of spacetime; they obey Maxwell's macroscopic equations when it comes to producing the $E, D, H, B$ fields, and they obey the E-L formulas for force-density and torque-density when it comes to calculating the effect of fields on these sources. Nowhere in the E-L formulation is it necessary to introduce models for $\rho_{\text{free}}$, $J_{\text{free}}$, $P$ and $M$. Simply stated, the sources are what they are in Nature, and they obey the aforementioned laws. Einstein and Laub do not necessarily see the magnetic dipoles as pairs of north and south monopoles; in fact, they scarcely dwell on the nature of these dipoles; their aim is simply to provide formulas for calculating the EM force and torque.

Experiments on hyperfine energy-splitting and neutron scattering involve one material particle residing inside (or crossing into) another material particle. The two overlapping particles generally attract or repel each other with equal and opposite forces, so that the net force on the pair of particles produced by their mutual interaction is zero. Should the E-L formulation be dismissed under such circumstances? The answer is no, because this formulation does *not* say anything at all about the internal structure of the two particles and cannot, therefore, be used to elucidate their mutual interaction. If the combined system of two overlapping particles can be envisioned as a single entity with its own (net) charge, current, polarization, and magnetization, then the E-L formulation gives the *net* force and torque acting on that composite particle. What the E-L formulas do *not* provide is the mutual attraction/repulsion of the (overlapping) particles for each other. In other words, Einstein and Laub do not allow *parsing* of the force/torque-density equations, Eqs.(7,8), in order to determine the resulting action on individual particles.

In contrast, the Lorentz formulation, based on its models for $p$ and $m$, suggests that the interaction energy for the overlapping pair of particles is given by a certain expression, e.g., $-m_1 \cdot m_2$ in the case of one magnetic dipole $m_1$ residing inside another magnetic dipole $m_2$. One is free, of course, to take the existence of this interaction energy as confirmation of the Lorentz force law at this level and under the circumstances. However, such direct matter-matter interactions are probably outside the domain of classical electrodynamics; moreover, they should exist not only for one dipole residing inside another, but also for every kind of elementary (or composite) particle which might overlap other such particles. Such interaction energies are generally needed to determine the scattering cross-sections of colliding particles, and typically go beyond the standard EM theory. In fact, Jackson's analysis in [22] is not so much concerned with the attractive/repulsive *forces* acting on individual (overlapping) particles as it is with the interaction energies and scattering cross-sections in the context of Schrödinger's equation.

If one is looking for evidence that magnetic dipoles are in fact made up of Amperian current loops (as opposed to Gilbertian pairs of north and south monopoles), then one need not look beyond the existence of orbital magnetic moments, or beyond the Lorentz transformation rules for $P$ and $M$ between inertial frames. However, looking at the problem in this way completely misses Einstein's subtle logic. Einstein and Laub in fact wrote two (back-to-back) papers in 1908. In their first paper [40], they derived formulas for the Lorentz transformation of $E, D, H, B$, $P$ and $M$ between different inertial frames. Knowing how $P$ and $M$ transform between inertial frames is tantamount to acknowledging that magnetic dipoles behave like Amperian current loops (at least as far as the Lorentz transformation of electric charge and current is concerned). Einstein and Laub knew this very well, as evidenced by the content of their first paper [40]. In their second paper [11], they postulated the now-famous Einstein-Laub formulas for EM force and torque densities. They seem to have been fully aware of their departure from the current-loop model of magnetic dipoles when they stated (in a footnote) that they would like to "*stick to the dual treatment of electric and magnetic phenomena for the sake of a simpler presentation.*"

Nearly a decade after Einstein had passed away, Shockley discovered "hidden momentum," which is perhaps rooted in Shockley's desire to see the Amperian current loop as a stable classical entity [23-25]. In contrast, Einstein, feeling that such stable loops may be impossible to concoct within classical physics, opted for fairly simple yet powerful formulas for EM force and torque, which avoided the notion of moving electric charges within a stable loop of current. This is a crucial difference between the Lorentz and E-L formulations of classical electrodynamics. It is not that Einstein doubted that, in certain contexts, magnetic dipoles behave similarly to loops of electrical current; and it is not that Einstein thought magnetic dipoles were made up of pairs of magnetic monopoles; rather, he appears to have strongly believed that EM force and torque exerted on these dipoles should be treated differently than in Lorentz's approach. Jackson's arguments in [22] provide yet another evidence for magnetic dipoles having similarities to Amperian current loops. Nevertheless, the stable nature of these loops, which are quantum mechanical in nature, precludes the *straightforward* application of the Lorentz force law to magnetic dipoles.

Perhaps the best approach to classical electrodynamics is to avoid models of electric and magnetic dipoles; instead, one should accept dipoles as equal partners with electrical charge and current, and use strict rules for calculating energy, momentum, angular momentum, force, and torque. These rules are simplest when ad hoc assumptions involving hidden energy, hidden momentum, corrections to force-density, corrections to torque-density, etc., are *not* introduced. A complete and consistent theory of electrodynamics can indeed be built around Maxwell's *macroscopic* equations, Poynting's theorem, Abraham's momentum, and the E-L force and torque laws. In every other respect the essential structure of the classical theory remains intact.

**Acknowledgement**. This paper originated in extensive discussions with Vladimir Hnizdo, David Griffiths, and John Weiner throughout 2014. The author is grateful to all for sharing their deep knowledge of classical electrodynamics and for posing insightful questions.